\documentclass[prb,twocolumn,floatfix,showpacs,groupedaddress]{revtex4}
\usepackage{graphicx,amsfonts,amssymb,hyperref,nicefrac,lipsum, verbatim, bbm}

\usepackage[fleqn]{amsmath}

\usepackage[version=3]{mhchem}

\usepackage{siunitx}  
\usepackage[x11names]{xcolor}
\newif\ifhyper
\hypertrue
\ifhyper
\hypersetup{
   citecolor = {green!60!black},
   colorlinks = {true}, 
   urlcolor = {blue} 
}
\fi

\newcommand{\be}{\begin{equation}}
\newcommand{\ee}{\end{equation}}
\newcommand{\beqa}{\begin{eqnarray}}
\newcommand{\eeqa}{\end{eqnarray}}
\newcommand{\ket} [1] {\vert #1 \rangle}
\newcommand{\bra} [1] {\langle #1 \vert}

\def\bra#1{\langle#1\vert}
\def\ket#1{\vert#1\rangle}

\def\Longarrow{\protect\@lra}
\def\@lra{\relbar\joinrel\relbar\joinrel\relbar\joinrel%
          \relbar\joinrel\rightarrow}

\begin{document}

\title{Topological magnon bands for magnonics}

\author{M.~Malki}
\email{maik.malki@tu-dortmund.de}
\affiliation{Lehrstuhl f\"ur Theoretische Physik 1, TU Dortmund, Germany}

\author{G.\ S.~Uhrig}
\affiliation{Lehrstuhl f\"ur Theoretische Physik 1, TU Dortmund, Germany}
\date{\rm\today}

\begin{abstract}
Topological excitations in magnetically ordered systems have 
attracted much attention lately. We report on topological 
magnon bands in ferromagnetic Shastry-Sutherland lattices whose
edge modes can be put to use in magnonic devices. The synergy of 
Dzyaloshinskii-Moriya interactions and geometrical frustration are responsible for the 
topologically nontrivial character. Using exact spin-wave theory, we determine the 
finite Chern numbers of the magnon bands which give rise to chiral 
edge states. The quadratic band crossing point vanishes due the present
anisotropies, and the system enters a topological phase. 
We calculate the thermal Hall conductivity as an experimental signature of
the topological phase. Different promising compounds are discussed as possible physical realizations of ferromagnetic Shastry-Sutherland lattices hosting the
required antisymmetric Dzyaloshinskii-Moriya interactions.
Routes to applications in magnonics are pointed out.
\end{abstract}


\maketitle

Topological phases \cite{hasan10, qi11} exist in both fermionic and bosonic systems and constitute a fast developing research area. Although the theoretical understanding of fermionic topological systems has made impressive progress, topological bosonic
excitations have gained considerable attention only in the past few years. 
Despite the increasing conceptual knowledge of topological matter, only very few materials have been identified with topological properties
compared to the large number of potential topological materials \cite{bradl17}. 
Even less is known about potential applications.
This is, in particular, true for topological bosonic signatures \cite{onose10}. Thus, it is a major challenge to theoretically predict and experimentally verify topological bosonic fingerprints
in order to move towards useful applications.

In the research of topological properties in condensed matter, the magnetic degrees 
of freedom have increased in importance. Magnetic data storage is already a 
ubiquitous everyday technology \cite{fert08}. Recently, magnetic spin waves,
so-called magnons, themselves are used to carry and to process information
which is called " magnonics " \cite{shind13, krugl10, demok12}. Adding topological
aspects the field of magnonics \cite{nakat17} considerably enhances the possibilities to build  efficient devices for which we will make a proposal in this paper.

The challenge in finding topological signatures in magnetically ordered 
spin systems are the small Dzyaloshinskii-Moriya (DM) interactions 
\cite{dzyal58,moriy60a} which induce only small Berry curvatures.
The size of the DM terms relative to the isotropic coupling is 
roughly as large as $|g-2|/2$, i.e., the deviation of the $g$ factor
from 2, because both result from spin-orbit coupling.
Thus, the DM terms are generically too small to induce detectable topological effects. In strongly frustrated systems,
however, the relative size of the DM terms can indeed be comparable to
the isotropic couplings \cite{splin16}.

Another issue is the localization of edge modes. Employing the wording of
semiconductor physics, one must distinguish direct (at fixed wave vector) and 
indirect gaps (allowing for changes in the wave vector). The existence of direct gaps throughout the Brillouin zone (BZ) is sufficient to separate bands so that their topological properties are well defined. But the vanishing of the indirect gap
generically implies that the edge states are not localized anymore, i.e., the bulk-boundary correspondence with respect to 
localization does not hold anymore \cite{malki18b}. 

In magnetic systems, three types of elementary excitations occur. Long-range
ordered magnets display magnons (or spin waves) \cite{bloch30}, valence-bond crystals mostly feature triplons \cite{schmi03c}, whereas quantum spin liquids 
may display fractional excitations \cite{morri09}, for instance, spinons 
\cite{fadde81}. For triplons, topological
behavior, i.e., non-zero Chern numbers \cite{thoul82}, has been predicted 
\cite{romha15,malki17a} and verified \cite{mccla17} in Shastry-Sutherland lattices and in spin ladders \cite{joshi17}.
For ferromagnetically ordered systems, topological magnons have been theoretically suggested in kagome lattices \cite{katsu10a,chisn15}, pyrochlore lattices 
\cite{zhang13}, and in honeycomb lattices \cite{owerr16, kim16}. For antiferromagnets,
they have been proposed in pyrochlore lattices \cite{li16c},
square, and cubic lattices exploiting the Aharonov-Casher effect \cite{nakat17b}.
In analogy to the quantum Hall effect \cite{klitz80}, the magnon Hall effect 
\cite{onose10} as well as the triplon Hall effect \cite{romha15} arise
since the topological Berry curvature acts analogously to a magnetic field.
So far, only the magnon Hall effect has been observed \cite{onose10, ideue12}. 
Topologically non-trivial spinons have been discussed in the 
Mott insulator \cite{pesin10, rache10} as well as in quantum spin liquids 
\cite{katsu10a, ruegg12, cho12}.

The Shastry-Sutherland model \cite{shast81b} is commonly studied with  antiferromagnetic couplings leading to triplon excitations \cite{miyah03}. Including DM interactions combined with a transverse magnetic field induces topological properties \cite{romha15,malki17a} where the magnetic field is also used as a control parameter to tune  topological phase transition. The Shastry-Sutherland lattice with purely ferromagnetic couplings serves also as a good platform but for topological magnon excitations. This is one of the two main objectives of this article;
the second one is to discuss compounds which are likely to realize
this model and to point out possible applications.

We show by exact spin-wave theory that the ferromagnetic Shastry-Sutherland model with DM couplings has topological bands with non-trivial Chern numbers. The
occurrence of a ferromagnetic ground state represents the spontaneous breaking of time-reversal symmetry. In combination with the DM interactions 
a finite Berry curvature is induced which may lead to finite
Chern numbers. The degeneracy at the quadratic band crossing point (QBCP) is lifted, and a gap opens. The expected topologically
protected edge states \cite{hatsu93} are retrieved in 
strip geometry \cite{malki17b}. In order to guide the experimental verification of the magnon Hall effect, we compute the thermal Hall effect as well.

Real materials are always three-dimensional ($3$D); so
we look for the ferromagnetic Shastry-Sutherland model realized in various layers 
of $3$D materials. If the interlayer coupling \cite{supplement} is not too strong,  the 
$3$D quantum Hall states can be considered as ensemble of layered two-dimensional ($2$D) quantum Hall states so that it is appropriate to investigate $2$D models. 

\begin{figure}[htb]
\centering
\includegraphics[width=0.9\columnwidth]{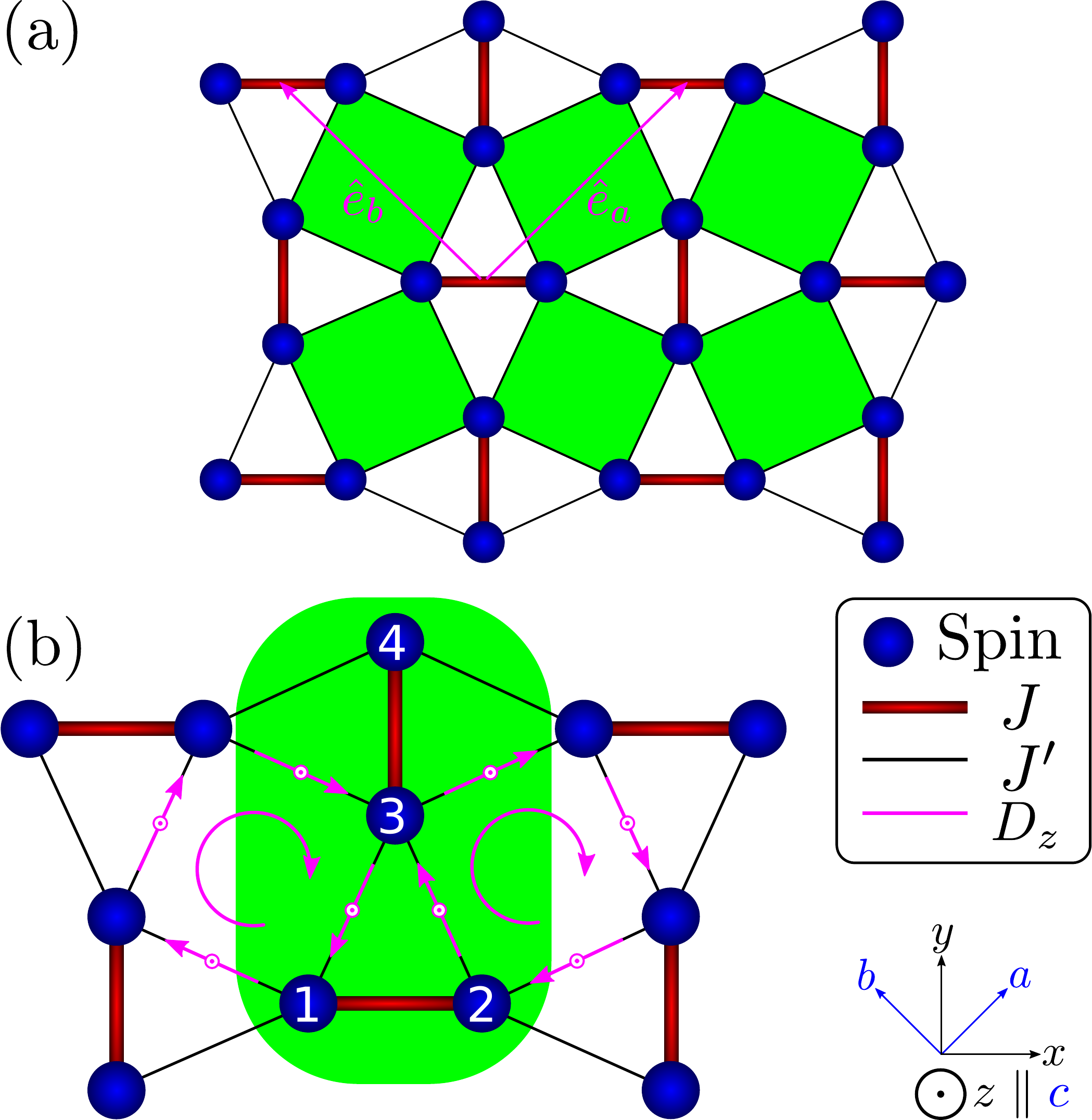} 
\caption{(a) Illustration of the $2$D Shastry-Sutherland lattice. The nearest neighbor couplings are shown as thick red lines, whereas the squares are highlighted in green. (b) The model studied comprises the Heisenberg couplings $J$ and $J'$ as well as the DM interaction $D_z$. The unit cell is highlighted in green. The sequence of the 
spin in the term $\mathbf{D}_z \cdot \mathbf{S}_i \times \mathbf{S}_j$ is shown by the arrow 
pointing from $i$ to $j$. The DM couplings follow a clockwise rotation, see the  circular arrows.}
\label{fig:lattice}
\end{figure}

Layers of the Shastry-Sutherland lattice are found in various insulating
magnetic materials since it is easily constructed from corner-sharing squares. The squares are not
aligned parallel or perpendicular to one other so that dimers are formed, see 
Fig.\ \ref{fig:lattice}(a). Due to the lack of inversion symmetry about the
midpoints of the bonds, DM interactions are possible and generically occur
from spin-orbit interactions. To reach large values of the DM couplings, it is
indicated to include atoms with large atomic numbers because large electron
velocities favor relativistic effects. 
Moreover, the couplings should be ferromagnetic so that it is indicated 
to avoid linear bonds which would favor antiferromagnetic superexchange according to the
Goodenough-Kanamori rules. Hence, the Shastry-Sutherland lattice
 depicted in Fig.\ \ref{fig:lattice} appears promising if superexchange
via larger subgroups does not occur (this is what happens in \ce{SrCu2(BO3)2}
\cite{miyah03}). The following materials appear to be particularly interesting: 
\ce{RE5Si4} or \ce{RE Si} 
\cite{roger06, spich01} (\ce{RE=Gd}, \ce{Dy}, \ce{Ho}, \ce{Er}, \ce{Y}). 
The compounds \ce{RE5Si4} have a \ce{Sm5Ge4}-type structure, and \ce{RESi} has a 
\ce{FeB}-type structure which both comprise planes of Shastry-Sutherland lattices. These compounds display a macroscopic magnetization $\mathbf{M}$
indicating dominant ferromagnetic couplings \cite{spich01,roger06}. In addition, the macroscopic magnetization clearly shows that one of the two degenerate ground states dominates, i.e., one domain prevails.

We compute the four bands from the unit cell with four sites 
shown in Fig.\ \ref{fig:lattice}(b) in green. The DM couplings of the Shastry-Sutherland lattice can be
 directed in plane or out of plane \cite{romha15}. Usually, however, the out-of-plane couplings dominate 
in 2D \cite{romha15,chern16}. In order to focus on a minimal model, we thus constrain the DM coupling to
 a uniform direction perpendicular to the plane $\mathbf{D} = D_z \hat{e}_z$ as shown in 
Fig.\ \ref{fig:lattice}(b). Obviously, this introduces a chiral orientation.
Single-ion anisotropy (SIA) $A^{\alpha \beta}$ ($\alpha, \beta \in \left\lbrace x, y, z \right\rbrace$) is typically present in ferromagnets with spins $S>1/2$. For the minimal model, we consider it to favor easy-axis alignment along the
$z$ axis $A^{zz} = A\ge 0$. The SIA and the DM coupling compete because
 the latter profits from tilts away from the $z$ axis.
A conservative estimate \cite{supplement} shows that for small SIA 
and DM coupling the SIA  wins and
the fully polarized state is generic.

The complete Hamiltonian of the minimal model consists of three parts
\begin{subequations}
\begin{align}
\mathcal{H} =& \mathcal{H}_{\mathrm{H}} + \mathcal{H}_{\mathrm{DM}} + 
\mathcal{H}_{\mathrm{SIA}}
\\
\mathcal{H}_{\mathrm{H}}  =& - J \sum_{\left\langle ij \right\rangle}  
\left[ \frac{1}{2} (S_i^+ S_j^- + S_i^- S_j^+) + S_i^z S_j^z \right] 
\nonumber \\
& -J' \! \!  \sum_{\left\langle \left\langle ij \right\rangle \right\rangle}\! 
\left[ \frac{1}{2} (S_i^+ S_j^- + S_i^- S_j^+) + S_i^z S_j^z \right] \! \! \! \! 
\\
\mathcal{H}_{\mathrm{DM}} =& - \frac{\mathrm{i} D_z}{2} 
\sum_{\left\langle \left\langle ij \right\rangle \right\rangle} (S_i^+ S_j^- - S_i^- S_j^+) 
\label{eq:dm}\\
\mathcal{H}_{\mathrm{SIA}} =& - A \sum_i (S_i^z)^2 , 
\label{eq:anisotropy}
\end{align}
\end{subequations} 
with ferromagnetic couplings $J, J' > 0$; $J$ serves as the energy unit henceforth. A pair of nearest neighbors and of next-nearest neighbors is denoted by 
${\left\langle ij \right\rangle}$ and by 
${\left\langle \left\langle ij \right\rangle \right\rangle}$, respectively.

\begin{figure}[htb]
\centering
\includegraphics[width=1.0\columnwidth]{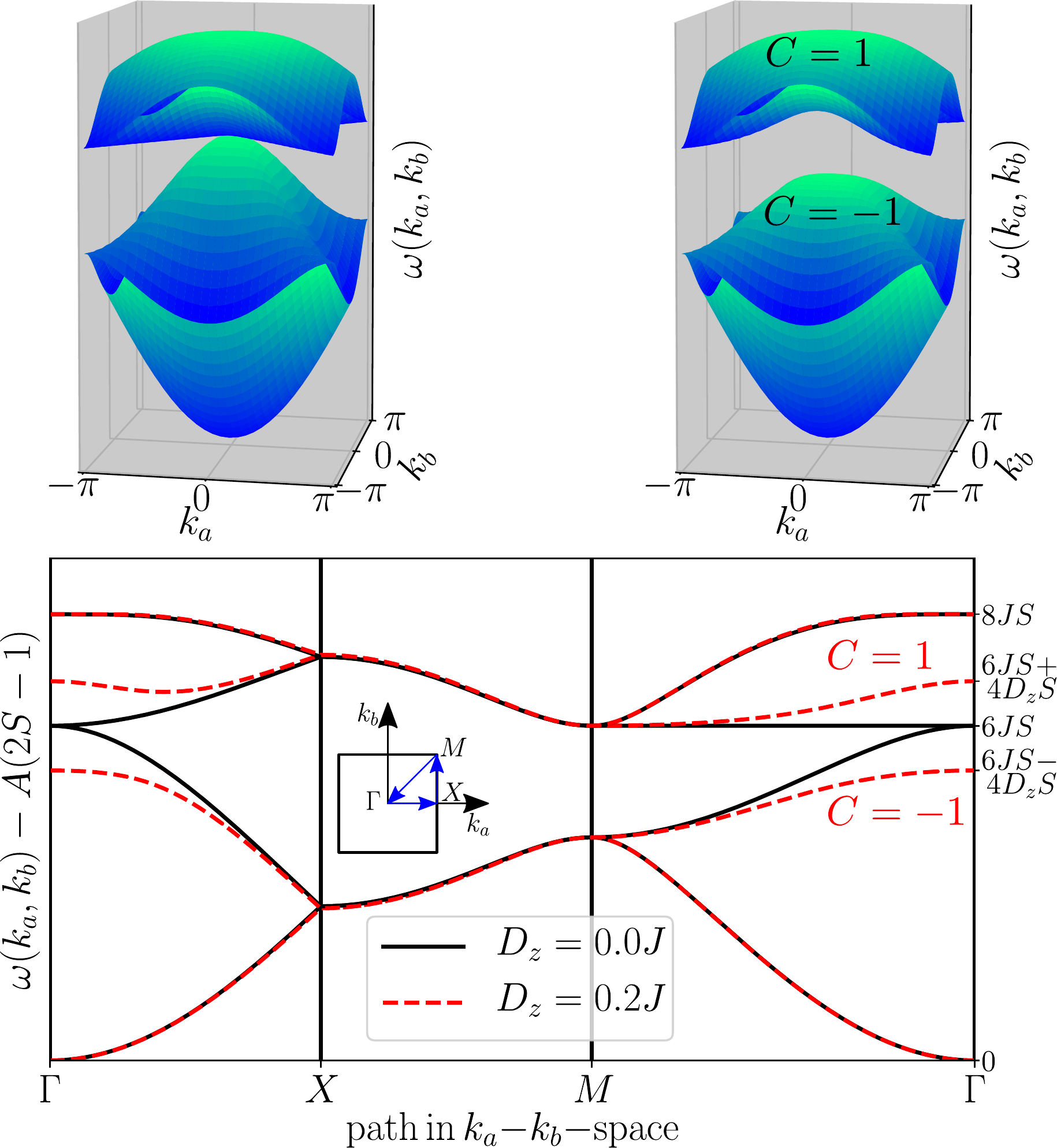} 
\caption{One-magnon dispersions for $J = J'$ for two values of
the DM coupling. The critical case is $D_z = 0$ (upper left panel and black lines
in the lower panel) where the QBCP at
the $\Gamma$ point is clearly visible. The degeneracy of the quadratic
bands is lifted for finite $D_z>0$ (upper right panel and red lines
in the lower panel) so that distinct bands appear
which show non-trivial topological Chern numbers $C = \pm 1$.}
\label{fig:dispersion}
\end{figure}

We use the Dyson-Maleev representation of the spin operators 
\cite{dyson56a, malee57} which is exact as long as a single magnon above
the fully polarized ground state is considered. But even for several
magnons, spin-wave theory is well justified due to the large spins involved
($S \approx 4-5$ for $\left\lbrace \ce{RE=Gd}, \ce{Dy}, \ce{Ho}, \ce{Er}, \ce{Y} \right\rbrace$). Note that large spins generically lead to large energy ranges with considerable gaps which are favorable for application.
The bilinear Hamiltonian in momentum space reads
\be
\mathcal{H} = \sum_{\mathbf{k}} 
\sum_{n m} b_{n, \mathbf{k}}^\dagger \mathcal{M}_{n m}^{\phantom{\dagger}} 
(\mathbf{k}) b_{m, \mathbf{k}}^{\phantom{\dagger}}
\ee
where $b_n^\dagger$ and $b_n^{\phantom{\dagger}}$ are the bosonic creation and annihilation operators at
 the site $n \in  \left\lbrace1, 2, 3, 4 \right\rbrace$, see \smash{Fig.\ 
 \ref{fig:lattice}(b).} The $4 \times 4$ Hamiltonian matrix is given by
\be
\underline{\underline{\mathcal{M}}}(\mathbf{k}) = \begin{pmatrix}
\underline{\underline{A}} & \underline{\underline{B}}(k_a, k_b) \\
\underline{\underline{B}}^{\dagger}(k_a, k_b) & \underline{\underline{A}}
\end{pmatrix} ,
\ee
with the $2 \times 2$ matrices,
\begin{subequations}
\begin{align}
\underline{\underline{A}} &= \begin{pmatrix}
JS  + 4 J'S + A (2S - 1) \quad \qquad -JS  \\
-JS  \quad  \qquad J S + 4 J' S + A (2S - 1)
\end{pmatrix} \\
\underline{\underline{B}} &= \begin{pmatrix}
- C^{\phantom{*}} (1 + e^{\mathrm{i} k_a}) & - C^* (e^{ \mathrm{i} k_a} + 
e^{ \mathrm{i} (k_a + k_b)} ) \\
- C^* (1 + e^{\mathrm{i} k_b}) & - C^{\phantom{*}} (e^{ \mathrm{i} k_b} + 
e^{ \mathrm{i} (k_a + k_b)} )
\end{pmatrix} \!\!\!\!\!\!
\end{align}
\end{subequations} 
where $C := S (J' + \mathrm{i} D_z )$. 
We set the lattice constant to unity so
that the wave vectors become dimensionless. Diagonalizing 
$\mathcal{M}(\mathbf{k})$ yields four distinct magnon bands $\mathcal{H} = 
\sum_{n, \mathbf{k}} \omega_n(\mathbf{k}) \tilde{b}^\dagger_{n, \mathbf{k}} 
\tilde{b}^{\phantom{\dagger}}_{n, \mathbf{k}}$ depicted  in Fig.\ \ref{fig:dispersion}.
The four bands come in pairs $p$ of two bands which are degenerate
on the boundary of the BZ. We strongly presume that this
degeneracy is linked to the composite point symmetry of the Shastry-Sutherland
lattice which consists of a vertical or horizontal translation
shifting vertical dimers to horizontal ones and vice versa combined with
a rotation by 90$^\circ$. But we did not find analytic proof.
The whole lattice is $C_4$ symmetric considering rotations about the centers
of the squares so that dispersions display the same symmetry. 

Ferromagnetic Heisenberg models without spin anisotropic couplings, such as SIA or DM coupling
display gapless Goldstone bosons \cite{golds61} with a quadratic dispersion at low energies 
at the $\Gamma$ point. As soon as the SIA $A$ is turned, on the continuous
spin rotation symmetry is no longer broken spontaneously but externally, and a finite spin gap 
$A(2S-1)$ appears; note the offset energy axis in the
lower panel of \smash{Fig.\ \ref{fig:dispersion}}. Spontaneously, 
the system chooses one of the two degenerate fully polarized ground states. 
This stabilizes the fully polarized ground state since it becomes
energetically isolated from the remaining spectrum.

For vanishing DM coupling, two magnon bands cross quadratically at the $\Gamma$ point
at finite energies. Hence, the model displays an unusual QBCP. Linear Dirac cones 
\cite{halda88b} or variants of them \cite{romha15} are more standard. 
Generically, one can assign a Berry phase of $\pi$
(or multiples of $\pi$) to them \cite{sun09b}. The QBCP is stable and can be interpreted as a pair of Dirac cones \cite{chong08} which are superimposed due to the $C_4$ symmetry \cite{sun09b}. As a result, a QBCP can have a Berry flux of $0$ or $\pm 2 \pi$. 
The QBCP can either be removed by breaking the $C_4$ symmetry which splits it
into an even number of Dirac cones or by lifting its degeneracy, e.g., by 
opening a gap leading to topologically non-trivial bands.
Turning on the DM interaction ($D_z \neq 0$) induces the latter scenario. 
But as shown in Fig.\ \ref{fig:dispersion}, 
the degeneracy of the upper pair of bands and of the lower pair of bands
at the boundary of the BZ persists so that no Chern number of a single band
can be defined. Hence, one defines the Chern number of subspaces by taking
the trace over the Berry curvature in each subspace \cite{soluy12, malki17a}  
which derives from the Berry phase of the determinants
of unitary transformations along closed paths \cite{uhrig91}. Denoting the Chern 
number of a pair of bands by $C^{(p)}$ where $p$ stands for `upper' or `lower'
one has
\begin{equation}
 C^{(p)} = \frac{1}{2\pi} \iint_{\rm BZ}  \sum_{n\in p}
\left[ F_{n, ab} ({\bf k})\; \right] \; {\rm d}k_a \; {\rm d}k_b \, ,
\end{equation}
where $F_{n,ab}$ is the Berry curvature of band $n$ defined by
\begin{subequations}
\begin{align}
 F_{n,ab} ({\bf k}) &= \frac{\partial A_{n,b} ({\bf k})}{\partial k_a} - 
\frac{\partial A_{n,a} ({\bf k})}{\partial k_b} \\
 \mathrm{with} \quad A_{n, \mu} (\mathbf{k}) &= \bra{\mathbf{k}, n} \nabla_{k_\mu}  
\ket{\mathbf{k}, n}\, .
\end{align}   
\end{subequations}

The numerical robust calculation of the Berry curvature is performed by discretization of 
the BZ \cite{fukui05} avoiding the eigenstates precisely at the boundaries of the BZ.
This is possible because the relevant curvature occurs in the vicinity of 
the $\Gamma$ point anyway. The calculated Chern numbers of the pairs of  magnon bands is
\smash{$C^{(\text{upper/lower})} = \pm 1$} as shown in Fig.\ \ref{fig:dispersion}. 
Changing the sign of the $D_z$ reverses the sign of the Chern numbers. 
The non-zero Chern numbers can be attributed to the complex hopping stemming from the
 DM coupling leading to fluxes of fictitious fields \cite{katsu10a, ideue12}. The direct gap
between both pairs of bands occurs at $\Gamma$ and is given by $8 D_z S$ as long as $4 D_z < J$. Otherwise, the direct gap is located at the $M$ point and takes the value $2 J S$. These relations highlight the importance of large spins and DM couplings for large gaps.

\begin{figure}[htb]
\centering
\includegraphics[width=1.0\columnwidth]{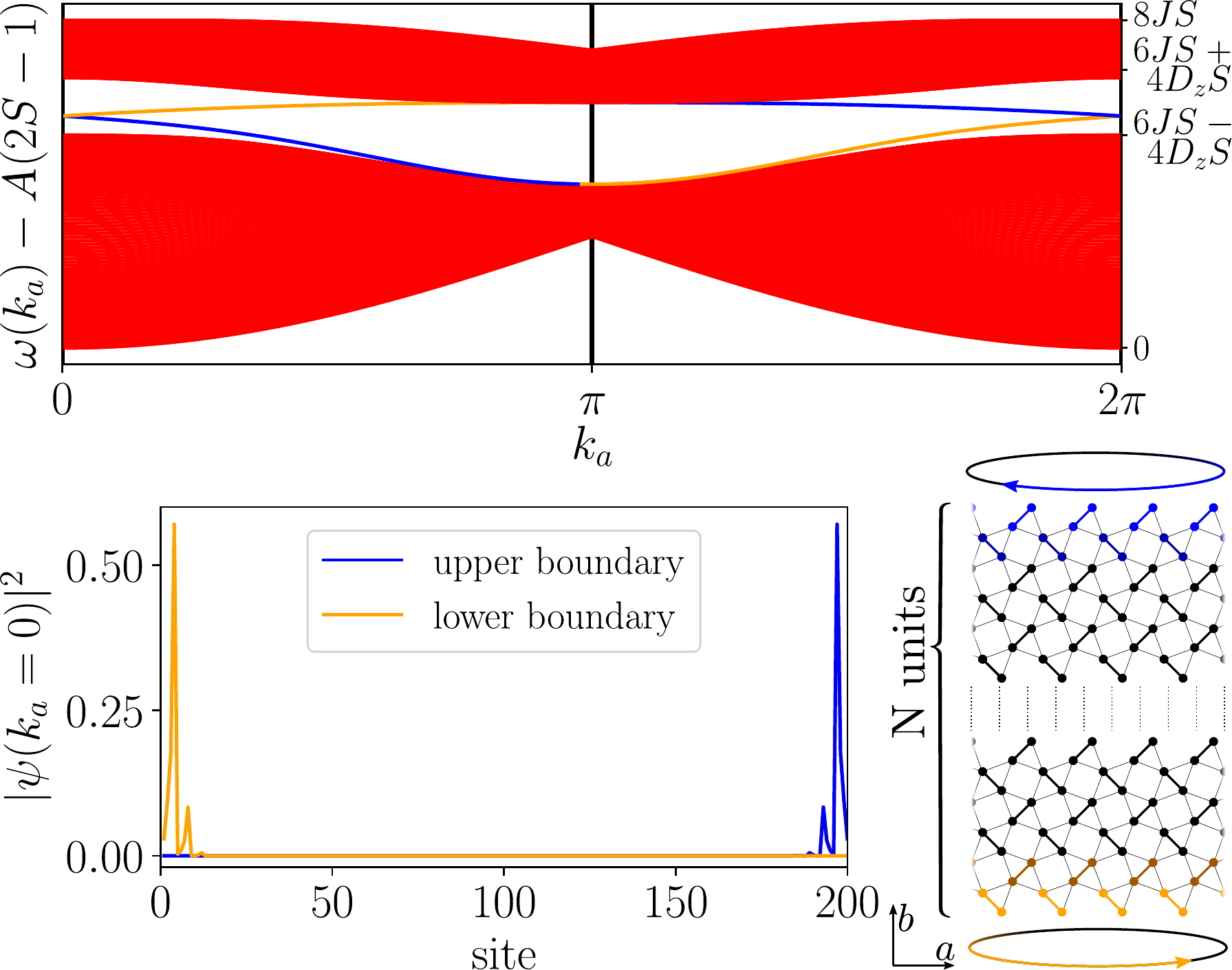} 
\caption{(Upper panel) Eigen energies of a strip geometry, see the lower right panel 
with $N = 50$, $J = J'$, \smash{$D_z = 0.2 J$}, and \smash{$A = 0.2 J$}. (Lower left panel) Probability density 
\smash{$|\psi(k_a = 0,r_b)|^2$} as a function of site $r_b$ of both edge modes;
the color of their curves corresponds to the color of the boundary sites in the lower right
panel.}
\label{fig:edge_states}
\end{figure}

According to the bulk-boundary correspondence \cite{hatsu93, thoul82}, the existence of nontrivial 
Chern numbers implies topologically protected edge states \cite{mook14}. For verification, we analyze a finite strip of $N = 50$ unit cells in the $b$ direction and periodic boundaries in the $a$ direction, see the 
lower right panel in Fig.\ \ref{fig:edge_states}. The energy eigen values 
as a function of the well-defined wave-vector $k_a$ are depicted in
 the upper panel of Fig.\ \ref{fig:edge_states}. 
One can easily see two chiral edge states moving right and left according
to the slope of their dispersion branches which connect the two continua shown in red.
Additionally, the lower left panel illustrates the localization of these modes
at the lower (yellow curve and sites) and upper (blue curve and sites) edge of the strip.

\begin{figure}[htb]
\centering
\includegraphics[width=1.0\columnwidth]{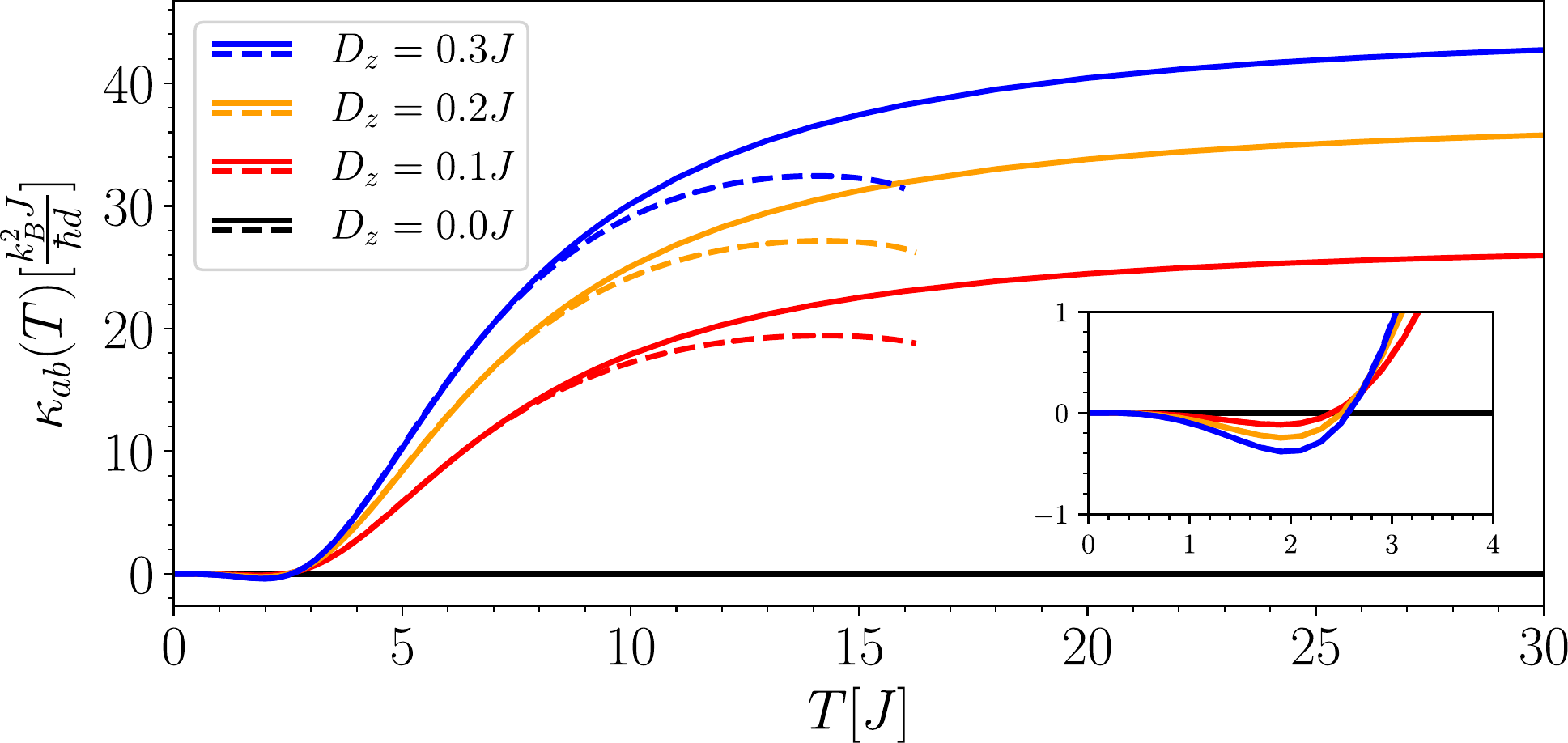} 
\caption{Thermal Hall conductivity $\kappa_{ab}$ as a function of temperature for various 
values of the DM coupling $D_z$ at \smash{$J = J',$} $A = 0.2 J$, and $S = 4$. }
\label{fig:hall}
\end{figure}

Next, we address possible experimental signatures. Since magnons do not carry charge, usual electric
conductivity measurements do not make sense. The thermal Hall effect offers a way to detect 
nontrivial Berry curvatures in real materials. The thermal Hall effect 
consists of a finite-temperature gradient perpendicular to a heat current. The expression for the transversal heat conductivity $\kappa_{ab}$ \cite{matsu11} is given by 
\begin{equation}
\label{eq:kappa}
\kappa_{ab} = - \frac{k_B^2 T}{\hbar} \sum_{n, \bf k} c_2(\rho_n) F_{n, ab} (\bf k)
\end{equation}
where we sum over all magnon bands and set $k_B = 1$  and $\hbar = 1$. The weight $c_2(\rho_n)$ is given by 
\begin{subequations}
\begin{align}
c_2(\rho) &= \left. \int_{\varepsilon_n}^{\infty} \mathrm{d} \varepsilon 
\left( \beta \varepsilon \right)^q \left( - \frac{\mathrm{d} \rho}{\mathrm{d} \varepsilon} \right) \right|_{\mu = 0} 
\\
 &= - 2 \, \mathrm{Li}_2 (-\rho) + \rho \ln^2(\rho^{-1} + 1) - \ln^2(\rho + 1) 
\nonumber \\
 &\phantom{=} + 2 \ln(\rho + 1) \ln(\rho^{-1} + 1) \, ,
\end{align}
\end{subequations}
where $\rho$ is the Bose-Einstein distribution $(\exp(\beta\omega)-1)^{-1}$ and $\mathrm{Li}_m$ is  the dilogarithm for $m=2$ (Spence's integral, in general). Equation \eqref{eq:kappa} clearly shows that the transversal heat conductivity $\kappa_{ab}$ depends directly on the Berry curvature, thus, representing
an ideal fingerprint of non-trivial topological properties. 
Figure \ref{fig:hall} displays the 
results of Eq.\ \eqref{eq:kappa} as a function of temperature for various values of $D_z$.
For the topological phase ($D_z \neq 0$), the conductivity first slightly decreases to negative values before it strongly increases as a function of temperature. For high temperatures, $\kappa_{ab}$ approaches
 a finite value. In comparison, the topologically trivial bands for $D_z = 0$ 
may have a finite curvature, but such that it cancels in the sum over the BZ so that  
$\kappa_{ab}$ vanishes.

Since the magnetization generally decreases with increasing temperature
untill eventually the ferromagnetic phase ceases to exist at $T_c$, $\kappa_{ab}$ 
should also decrease until it disappears at $T_c$. By improving the calculations by 
applying self-consistent spin-wave theory, the signature starts to decrease for higher 
temperatures before no self-consistent solution is found anymore as depicted by the 
dashed lines \cite{supplement}.

In conclusion, a finite thermal Hall conductivity $\kappa_{ab}$ can serve as a smoking 
gun signature in experiments to verify topological properties of a material under 
study. For large signals, we suggest the experimental preparation of a single domain crystal in order to avoid cancellation effects.

In view of the above findings, we suggest characterizing the magnetic properties of
the putative realizations \cite{spich01,roger06} of ferromagnetic Shastry-Sutherland 
lattices in detail,
for instance, by inelastic neutron scattering. This will help to determine the 
relevant microscopic model which, in turn, will render the calculation of the Berry 
curvature
possible. In parallel, measurements of the thermal Hall conductivity can
provide evidence \cite{onose10, ideue12} for finite Berry curvatures.

The intriguing next step towards an application will be to tailor the edges
of strip by decorating them similar to what has been proposed and computed
for fermionic models \cite{uhrig16,malki17b}. In this way, largely different group 
velocities can be achieved depending on the direction in which signals of packets
of magnons travel. The key is to structure the upper and the lower
boundaries of a strip in a different manner so that the group velocity 
of the right- and of the left-moving packet is very different.
Ideally, the group velocities should be tunable by
moderate changes of the model controlled by external parameters,
such as magnetic fields or pressure. The realization of this
phenomenon will pave the way for fascinating devices in magnonics, such
as delay lines and interference devices.


\acknowledgments
This paper was also supported by the Deutsche Forschungsgemeinschaft and the Russian Foundation of Basic Research in the International Collaborative Research Center TRR 160.
MM gratefully acknowledges financial support by the Studienstiftung des deutschen Volkes. GSU thanks the School of Physics of the University of New South Wales for its hospitality during the preparation of paper and the Heinrich-Hertz Stiftung for financial support of this stay. We thank C.\ H.\ Redder and O.\ P.\ Sushkov for useful discussions. 

%

\clearpage
\pagebreak

\setcounter{equation}{0}
\setcounter{figure}{0}
\setcounter{table}{0}
\setcounter{page}{1}
\makeatletter
\renewcommand{\theequation}{S\arabic{equation}}
\renewcommand{\thefigure}{S\arabic{figure}}
\renewcommand{\bibnumfmt}[1]{[S#1]}
\renewcommand{\citenumfont}[1]{S#1}

\vspace*{0.5cm}

\onecolumngrid
\begin{center}
\textbf{\large  Supplemental Material for "Topological magnon bands for magnonics"}
\end{center}
\twocolumngrid

\section*{Estimate of the competition between the single-ion anisotropy 
and the Dzyaloshinskii-Moriya interaction}

Here we estimate up to which value of the Dzyaloshinskii-Moriya (DM) interaction the collinear, fully polarized ferromagnetic order favored by the
single-ion anisotropy (SIA) represents the ground state of the model. 
To obtain such an estimate we study two 
next-nearest neighbor spins coupled by $J'$ as classical vectors of length $S$ 
with polar angles $\theta_1$ and $\theta_2$ and relative azimuthal angle $\varphi:=\varphi_1-\varphi_2$ which takes the value $\tan\varphi=d:=D_z/J'$ at the energy minimum $E$ 
\begin{equation}
2E/(J' S^2) = -a(1+xy)/2 -x-y-|x-y|\sqrt{1+d^2}
\end{equation}
where $a:=A/J'$, $x:=\cos(\theta_1+\theta_2)$, and $y:=\cos(\theta_1-\theta_2)$
with $|x|,|y|\le1$. 
The SIA term $a$ is split into four parts $a/4$ because each site has four $J'$ bonds. As long as $1+a/2\ge \sqrt{1+d^2}$  full polarization is optimum, i.e., a canted state can occur 
for $d\ge\sqrt{a+a^2/4}$ only, which is a conservative
estimate because the effects of $J$, of quantum fluctuations, and of the geometric constraints  
in the lattice are not included. Hence, for small SIA and DM coupling the SIA  wins and
the fully polarized state is generic.

\section*{Effects of interlayer couplings}

In absence of detailed information about the structure and the magnetic couplings
in the discussed three-dimensional (3D) materials we elucidate that the effect of 
weak interlayer couplings do not destroy the topological properties put
forward in the main text.

To this end, we assume that the system consists of stacked parallel planes present where each 
plane realizes a two-dimensional (2D)
 ferromagnetic Shastry-Sutherland model. The distinct planes are connected by 
a perpendicular interlayer coupling $J_c$. Since there is no detailed data on magnetic exchange 
paths for the proposed classes of materials \cite{roger06, spich01} we restrict the calculations
 to vertical couplings between the layers because they usually have the largest impact. As a result, the Hamiltonian Eq.\ (1) is extended by the additional term
\begin{equation}
\mathcal{H}_\mathrm{inter} = - J_c \sum_{\left\langle ij \right\rangle}  
\left[ \frac{1}{2} (S_i^+ S_j^- + S_i^- S_j^+) + S_i^z S_j^z \right] 
\end{equation}
with the ferromagnetic coupling $J_c > 0$. The notation $\left\langle ij \right\rangle$ 
indicates a coupling between nearest neighbors from adjacent layers. We apply the Dyson-Maleev representation 
\cite{dyson56a, malee57} and the Fourier transform to calculate the dispersion of the 
bosonic one-particle excitations within spin wave theory which is exact for the 
excitations above the fully polarized ferromagnetic state.
In this way, the interlayer term $\mathcal{H}_\mathrm{inter}$ results
to an additional  $4 \times 4$-matrix  given by $4 J_c S \sin^2(k_c^2/2) \mathbbm{1}$, i.e.,
proportional to the identity matrix.

From this we conclude that the ground states remains fully polarized since the finite 
spin gap remains at its value $A(2S - 1)$ created by the SIA. 
The topological properties in the bulk remain  the same as well because terms proportional 
to the identity matrix obviously do not change the eigen states. 
Since only the eigen states determine the topological properties the same Chern number
will ensue, regardless of the strength of $J_c$, at each value of $k_c$. 

Concomitantly, we find edge states in the corresponding strip geometry for 
arbitrary $k_c$. The resulting dispersion is the same as shown in Fig. 3 (in the 
main text) with an 
additional overall shift proportional to $4 J_c S \sin^2(k_c^2/2)$. The 
localization of the 3D edge states is identical due to the one of the 2D eigen states. 
The set of all localized edge states depending on $k_a$ and $k_c$ represent surface states.

In conclusion, the investigation of additional 3D perpendicular interlayer coupling shows that 
the topological edge modes persist and are not altered  as long as the fully polarized ground state is preserved. Hence, for weak interlayer coupling the topological properties found in the 
two-dimensional model also hold  in three dimensions and thus the proposed materials are
good candidates to search for realizations.

\section*{Self-consistent spin wave theory} 

The temperature dependency of the static calculation of the transversal heat 
conductivity $\kappa_{ab}$ stems only from the weight $c_2(\rho_n)$ and its contribution
to $\kappa_{ab}$ which is proportional to the Berry curvature $F_{n, ab}(\mathbf{k})$.
This curvature, however, is independent of temperature. In 
order to make quantitative statements, it is appropriate to improve the results
for finite temperatures, i.e., to include partly the effects of 
finite $T$. 

Here we use the Dyson-Maleev representation of the spin operators which leads to an 
exact description at quartic level and is given by 
\begin{align}
S_i^+ &= \sqrt{2S} (b_i^{\phantom{\dagger}} - \frac{1}{2S} b_i^\dagger b_i^{\phantom{\dagger}} b_i^{\phantom{\dagger}})  \\
S_i^- &= \sqrt{2S} \phantom{(} b_i^\dagger \\
S_i^z &= b_i^\dagger b_i^{\phantom{\dagger}} - S \, .
\end{align}

The complete Hamiltonian in the bosonic representation is then described by

\begin{subequations}
\begin{align}
\mathcal{H} =& \mathcal{H}_{\mathrm{H}} + \mathcal{H}_{\mathrm{DM}} + 
\mathcal{H}_{\mathrm{SIA}}
\\
\mathcal{H}_{\mathrm{H}}  =& - J \sum_{\left\langle ij \right\rangle}  
 S (b_i^\dagger b_j^{\phantom{\dagger}} + b_j^\dagger b_i^{\phantom{\dagger}} - b_i^\dagger b_i^{\phantom{\dagger}} - b_j^\dagger b_j^{\phantom{\dagger}}) \\
 &+ J \sum_{\left\langle ij \right\rangle} \frac{1}{2}(b_i^\dagger b_i^\dagger b_i^{\phantom{\dagger}} b_j^{\phantom{\dagger}} + b_j^\dagger b_j^\dagger b_j^{\phantom{\dagger}} b_i^{\phantom{\dagger}}) - b_i^\dagger b_j^\dagger b_i^{\phantom{\dagger}} b_j^{\phantom{\dagger}} \! \! \! \! \nonumber  \\
& -J' \! \!  \sum_{\left\langle \left\langle ij \right\rangle \right\rangle}\! S (b_i^\dagger b_j^{\phantom{\dagger}} + b_j^\dagger b_i^{\phantom{\dagger}} - b_i^\dagger b_i^{\phantom{\dagger}} - b_j^\dagger b_j^{\phantom{\dagger}}) \\
 &+ J' \! \! \sum_{\left\langle \left\langle ij \right\rangle \right\rangle} \frac{1}{2}(b_i^\dagger b_i^\dagger b_i^{\phantom{\dagger}} b_j^{\phantom{\dagger}} + b_j^\dagger b_j^\dagger b_j^{\phantom{\dagger}} b_i^{\phantom{\dagger}}) - b_i^\dagger b_j^\dagger b_i^{\phantom{\dagger}} b_j^{\phantom{\dagger}} \! \! \! \! \nonumber  \\
\mathcal{H}_{\mathrm{DM}} =& - \mathrm{i} D_z 
\sum_{\left\langle \left\langle ij \right\rangle \right\rangle} S(b_i^\dagger b_j^{\phantom{\dagger}} + b_j^\dagger b_i^{\phantom{\dagger}}) \nonumber \\
 &+ \frac{\mathrm{i} D_z}{2} \sum_{\left\langle \left\langle ij \right\rangle \right\rangle}(b_i^\dagger b_i^\dagger b_i^{\phantom{\dagger}} b_j^{\phantom{\dagger}} + b_j^\dagger b_j^\dagger b_j^{\phantom{\dagger}} b_i^{\phantom{\dagger}}) \\
\mathcal{H}_{\mathrm{SIA}} =& - A \sum_i (2S - 1) b_i^\dagger b_i^{\phantom{\dagger}} +  b_i^\dagger b_i^\dagger b_i^{\phantom{\dagger}} b_i^{\phantom{\dagger}} \, ,
\end{align}
\end{subequations}

\noindent where we neglected all constant terms. Applying a mean field decoupling 
reduces the quartic terms into bilinear terms. For this purpose we introduce the 
expectation values
\begin{subequations}
\begin{align}
n &= \left\langle b_i^{\dagger} b_i^{\phantom{\dagger}} \right\rangle \in \mathbb{R} \\
a &= \left\langle b_i^{\dagger} b_j^{\phantom{\dagger}} \right\rangle \in \mathbb{R} \quad \mathrm{for} \left\langle ij \right\rangle\\
c &= \left\langle b_i^{\dagger} b_j^{\phantom{\dagger}} \right\rangle \in \mathbb{C} \quad \mathrm{for} \left\langle \left\langle ij \right\rangle \right\rangle \, ,
\end{align}
\end{subequations}

\noindent where $a$ corresponds to nearest neighbors and $c$ to next-nearest neighbors. 
The Fourier transformation of the mean field Hamiltonian yields
\begin{equation}
\mathcal{H} = \sum_{\mathbf{k}} 
\sum_{n m} b_{n, \mathbf{k}}^\dagger \mathcal{M}_{n m}^{\phantom{\dagger}} 
(\mathbf{k}, n, a, c) \, b_{m, \mathbf{k}}^{\phantom{\dagger}}
\end{equation}

\noindent with the bosonic creation $b_n^\dagger$ and annihilation operators 
$b_n^{\phantom{\dagger}}$ at the site $n \in  \left\lbrace1, 2, 3, 4 \right\rbrace$. The Hamiltonian becomes implicitly temperature dependent since the expectation values are depending on temperature. The $4 \times 4$ Hamilton matrix reads 
\begin{equation}
\underline{\underline{\mathcal{M}}}(\mathbf{k}) = \begin{pmatrix}
\underline{\underline{A}} & \underline{\underline{B}}(k_a, k_b) \\
\underline{\underline{B}}^{\dagger}(k_a, k_b) & \underline{\underline{A}}
\end{pmatrix}
\end{equation}

with the $2 \times 2$ matrices

\begin{align}
\underline{\underline{A}} = 
\begin{pmatrix} 
A_{11} & A_{12} \\
A_{21} & A_{22} \\
\end{pmatrix} \qquad
\underline{\underline{B}} = \begin{pmatrix}
B_{11} & B_{12} \\
B_{21} & B_{22} \\
\end{pmatrix} \!\!\!\!\!\!
\end{align}
\begin{widetext}
\begin{subequations}
\begin{align}
A_{11} &= A_{22} = J(S - n + a)  + 4 J' (S - n + \mathrm{Re}(c)) + A (2S -4n- 1) + 4 D_z \mathrm{Im}(c) \\
A_{12} &= A_{21} = -J(S - n + a)  \\
B_{11} &= - J' (S - n + c^*) - J' (S - n + c) e^{\mathrm{i} k_a} - \mathrm{i} D_z (S - n) (1 + e^{\mathrm{i} k_a}) \\
B_{21} &= - J' (S - n + c^*) - J' (S - n + c) e^{\mathrm{i} k_b} + \mathrm{i} D_z (S - n) (1 + e^{\mathrm{i} k_b}) \\
B_{12} &= - J' (S - n + c)e^{\mathrm{i} k_a} - J' (S - n + c^*)e^{\mathrm{i} (k_a + k_b)} + \mathrm{i} D_z (S - n) (e^{\mathrm{i} k_a} + e^{ \mathrm{i} (k_a + k_b)} ) \\
B_{22} &= - J' (S - n + c)e^{\mathrm{i} k_b} - J' (S - n + c^*)e^{\mathrm{i} (k_a + k_b)} - \mathrm{i} D_z (S - n) (e^{\mathrm{i} k_b} + e^{ \mathrm{i} (k_a + k_b)} )
\end{align}
\end{subequations}
\end{widetext}

\begin{figure}
\centering
\includegraphics[width=1\columnwidth]{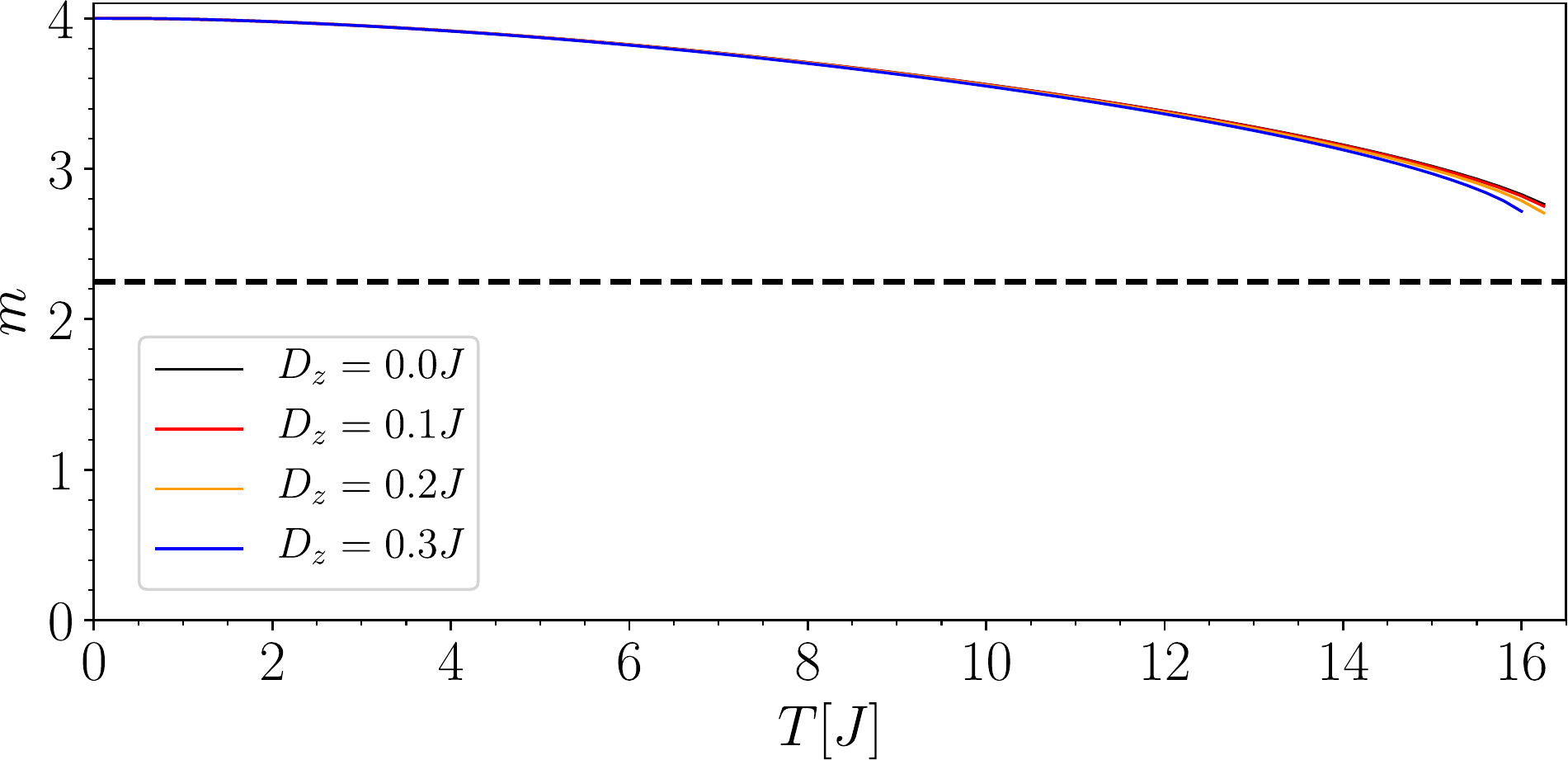} 
\caption{Magnetization as a function of temperature $T$ for various $D_z$ at $J=J', A = 0.2 J,$ and $S = 4$. The 
horizontal dashed line indicates the value at which the spin gap closes.}
\label{fig:magnetization}
\end{figure}

By expressing the expectation values using the Bose-Einstein distribution we are 
able to determine self-consistently the renormalized dispersion and the corresponding 
magnetization $m$ at a specific temperature. The magnetization is given by the simple 
relation \smash{$m = S - n$}. The renormalized spin gap $\Delta$ is purely determined by the SIA being 
given by 
\begin{equation}
\Delta = A (2S - 4n - 1).
\end{equation} 
\noindent Obviously, the spin gap closes before the magnetization 
vanishes, so that in this approximation a Curie temperature cannot be determined. 
The spin gap closes for $2S - 4n - 1 = 0$. For the magnetization this implies that the spin gap closes if the magnetization reaches the value $m = (2S + 1)/4$ as indicated by the horizontal dashed line in \smash{Fig. \ref{fig:magnetization}}.

The self-consistently calculated magnetization shows the unexpected problem that no solution 
can be found even before the spin gap closes or the magnetization vanishes. It appears that 
the phase transition from the ordered phase induced by the SIA to the disordered phase cannot 
be captured by spin wave theory. This issue deserves further investigations, but 
it is beyond the scope of the present article.

\newpage

\end{document}